\documentclass[twocolumn,showpacs,preprintnumbers,amsmath,amssymb]{revtex4}

\usepackage{graphicx}
\usepackage{dcolumn}
\usepackage{bm}
\usepackage{amssymb}

\begin{document}
\title{Is the photon paramagnetic?}
\author{ H. P\'erez Rojas and E. Rodriguez Querts}
\affiliation{Instituto de Cibernetica, Matematica y Fisica, Calle E
309, Vedado, Ciudad Habana, Cuba.\\}
\date{\today}

\begin{abstract}
A photon exhibits a tiny anomalous magnetic
 moment $\mu_{\gamma}$ due to its interaction  with an external constant magnetic field in vacuum through
 the  virtual electron-positron  background. It is
 paramagnetic ($\mu_{\gamma}>0$) in the whole region of transparency, i.e. below the first threshold energy
 for pair creation and has a maximum near this threshold.
The photon magnetic moment is different for eigenmodes polarized
along and perpendicular to the magnetic
 field.  Explicit expressions are given for $\mu_{\gamma}$ for the cases of photon
energies  smaller and closer to the first pair creation threshold.
The region beyond the first threshold is briefly discussed.
\end{abstract}



\maketitle \section{Introduction}
 In recent years, due to the development of high power lasers and ion accelerators, the problem
 of pair creation in strong
 external electromagnetic fields has attracted
 the interest of several researchers both in experimental and theoretical aspects
 (see for instance \cite{Muller1}-\cite{Muller3} and references therein).
 In this connection, it is interesting to turn our attention to the study of some
  elementary particle properties  arisen from radiative
 corrections in strong external electromagnetic fields. For instance, in
 analogy to the anomalous magnetic moment
$\mu^\prime=\alpha\mu_B/2\pi$ (being $\mu_B=e/2m$ the Bohr magneton)
for electrons shown by Schwinger \cite{Schwinger} as due to their
interaction with the virtual photon background through its
self-energy,  we want to show in the present paper that a similar
effect  exist for photons. Due to the magnetic properties of the
photon self-energy, a photon anomalous magnetic moment
$\mu_{\gamma}>0$  arises, which is paramagnetic in the region of
transparency (which is the region of momentum space where the photon
self-energy, and in consequence, its frequency, is real), and has a
maximum value for $\omega$ close to the first threshold for pair
creation $\omega=2m$. The photon magnetic moment vanishes only when
its momentum $\textbf{k}$ is  parallel to the magnetic field
$\textbf{B}$.

 The photon magnetic properties are due to the
 dependence of $\omega$   on $B$ expressed by the photon dispersion
 equation dependence on the
self-energy tensor $\Pi_{\mu\nu}(x,x^{\prime\prime}\vert A^{ext})$.
For an observable photon the quantity $\mu_{\gamma}=-\partial
\omega/\partial B$ can be obtained analytically from the expression
for $\Pi_{\mu\nu}(x,x^{\prime\prime}\vert A^{ext})$. In terms of the
$e^{\pm}$ Green functions $G(x,x')$ in presence of the external
field $B$ it is $\Pi_{\mu\nu}(x,x^{\prime\prime}\vert
A^{ext})=\alpha \int Tr
\gamma_{\mu}G(x,z)\gamma_{\nu}G(z,x^{\prime\prime})d^4 z$ and in the
one-loop approximation it was calculated by Batalin and Shabad in
the Schwinger proper time representation \cite{batalin}. Such
expression contains the sum over all Landau numbers and spin quantum
numbers of the virtual $e^{\pm}$ pairs. From the dispersion
equations for the eigenmodes resulting from its  diagonalization,
 it was found in \cite{shabad1,shabad2} that the photon suffers a strong deviation from the
light cone curve at frequencies near the pair creation energy
thresholds, indicating that the photon dynamics in the external
magnetic field is strongly influenced by the virtual $e^{\pm}$
pairs, showing a behavior similar to a massive particle. We
emphasize here that the photon propagating in magnetized vacuum
behaves in a similar way to a polariton, which in condensed matter
are quasiparticles resulting from strong coupling of a photon with a
dipole-carrying excitation.

The contribution from the electron anomalous magnetic moment to the
photon anomalous magnetic moment would appear in the two loop
approximation, of order $\alpha^2$ with regard to the radiation
field, and not in the present one-loop term, proportional to
$\alpha$.

We start in the first Section by recalling some basic features of
the quantum relativistic electron  dynamics in the magnetic field
$B$ and on the photon eigenmodes kinematics and dynamics.  In the
second we obtain the general expression for the photon magnetic
moment in terms of the self-energy eigenvalues. In the third
Section, we obtain the magnetic moment in the small frequency limit,
which leads actually to expressions valid up to frequencies close to
the threshold limit $2m$. The fourth Section is devoted to obtain
the approximate expressions for the photon magnetic moment at
frequencies near and below the pair creation thresholds, where it
has a maximum value. In the last Section a resume is made of the
results obtained in the paper, which correspond to the so-called
region of transparency, and mention is made about some features
characterizing the region of absorption, beyond the first threshold
for pair creation.

\section{Some basic results}
If the constant uniform magnetic field $B$ is along the $x_3$ axis,
it breaks the space symmetry so that for electrons and positrons
($e^{\pm}$) physical quantities are invariant only under rotations
around $x_3$ or displacements along it. Angular momentum and spin
components $J_3$,$L_3$,$s_3$ as well as linear momentum $p_3$ are
conserved. By using units $\hbar=c=1$, the energy eigenvalues for
$e^{\pm}$ are $E_{n,p_3}=\sqrt{p_3^2+m^2+ eB(2n+1+s_3)}$ where
$s_3=\pm 1$ are spin eigenvalues along $x_3$ and $n=0,1,2..$ are the
Landau quantum numbers \cite{Johnson}. In other words, in presence
of $B$, the transverse squared energy $E_{n,p_3}^2-p_3^2$ is
quantized by integer multiples of $eB$. For the ground sate $n=0$,
$s=-1$, the integer is zero. Quantum states degeneracy with regard
spin is expressed by a term $\alpha_n=2-\delta_{0n}$, whereas
degeneracy with regard to orbit' center coordinates leads to a
factor $eB$, The quantity $1/eB$ characterizes the spread of the
$e^{\pm}$ spinor wavefunctions in the plane orthogonal to $B$. Due
to the explicit symmetry breaking, the four momentum operator acting
on the vacuum state does not have a vanishing four-vector
eigenvalue, $P_{\mu}|0,B> \neq 0$. The $e^{\pm}$ quantum vacuum
energy density is given by $\Omega_{EH}=-eB\sum_{n=0}^\infty
\alpha_n \int dp_3 E_{n,p_3}$, After removing divergences it gives
the well-known Euler-Heisenberg expression $\Omega_{EH}=\frac{\alpha
B^2}{8\pi^2}\int_0^{\infty}e^{-B_c x/B}\left[\frac{coth x}{x}
-\frac{1}{x^2}-\frac{1}{3}\right]\frac{d x}{x}$ which is an even
function of $B$ and $B_c$, where $B_c=m^2/e\simeq 4.4\times
10^{13}$G is the Schwinger critical field. The magnetized vacuum is
paramagnetic ${\cal M}_V=-\partial \Omega_{EH}/\partial B>0$ and is
an odd function of $B$ \cite{Elizabeth}. For $B<<B_c$ it is ${\cal
M}_V=\frac{2\alpha}{45 \pi}\frac{B^{3}}{B_c^2}$, where $\alpha$ the
fine structure constant.

  The diagonalization of the photon self-energy tensor leads to
the equations \cite{shabad1}
\begin{equation}
 \Pi_{\mu
\nu}a^{(i)}_{\nu}=\kappa_{i}a^{(i)}_{\mu},
\end{equation}
 having
three non vanishing eigenvalues
 and three eigenvectors for $i=1,2,3$, corresponding to three photon propagation modes. One additional eigenvector is
the photon four momentum vector $k_{\nu}$ whose eigenvalue is
$\kappa_{4}=0$. The first three eigenvectors
\begin{eqnarray} \nonumber
    a^{1}_\mu= k^2 F^2_{\mu \lambda}k^\lambda-k_\mu (kF^2 k),
     \\
    a^{2}_\mu=F^{*}_{\mu \lambda}k^\lambda,\hspace{.3cm} a^{3}_\mu=F_{\mu
\lambda}k^\lambda,
\end{eqnarray}
satisfy the four dimensional transversality condition
$a^{(1,2,3)}_{\mu}k_{\mu}=0$. Here $k_\mu$ is the photon
four-momentum, $F_{\mu \nu}=\partial_\mu A_\nu-\partial_\nu A_\mu$
and $F^{\mu \nu*}=\frac{1}{2}\epsilon^{\mu \nu \rho \kappa}F_{\rho
\kappa}$ are the external electromagnetic field tensor and its dual
pseudotensor, respectively.
The vectors ${\bf k}_{\perp}$ and ${\bf k}_{\parallel}$ are the
components of $\textbf{k}$ across and along $\bf B$. In what follows
$k^2=z_1 + z_2$ where $z_{1,2}$ are invariant variables defined by
\begin{equation}
   z_1=\frac{kF^{*2}k}{2\mathfrak{F}}=k_\parallel^2-\omega^2,
   \hspace{1cm}z_2=-\frac{kF^{2}k}{2\mathfrak{F}}=k_\perp^2.
\end{equation}

In reference frames which are at rest or moving parallel to
$\textbf{B}$ we define $\textbf{n}_\perp=\textbf{k}_\perp/k_\perp$
and $\textbf{n}_\parallel=\textbf{k}_\parallel/ k_\parallel$ as the
transverse and parallel unit vectors respectively.

By considering $a^{(i)}_\mu (x)$  as the electromagnetic four vector
describing these eigenmodes, its electric and magnetic fields ${\bf
e^{(i)}}=-\frac{\partial }{\partial x_0}\vec{a}^{(i)}-\frac{\partial
}{\partial {\bf x}}a^{(i)}_0$, ${\bf
h}^{(i)}=\nabla\times\vec{a}^{(i)}$ are obtained in \cite{shabad1}.
 It is easy to see
that the mode $i=3$ is a transverse plane polarized wave whose
electric unit vector is $\textbf{e}^{(3)}= (\textbf{n}_\perp\times
\textbf{n}_\parallel)$ orthogonal to the plane ($\textbf{B},
\textbf{k}$). For  $\bf k \perp \bf B$, $\textrm{a}^{(1)}_\mu$ is
longitudinal,  polarized along $\textbf{e}^{(1)}=\textbf{n}_\perp$
and it is a non physical mode, whereas $\textrm{a}^{(2)}_\mu$ is
transverse, since $\textbf{e}^{(2)}=\textbf{n}_\parallel$. Thus, in
that case modes $\textrm{a}^{(2,3)}_\mu$ are superposition of waves
of opposite helicity. For
 $\bf k \parallel \bf B$, the mode $\textrm{a}^{(2)}_\mu$
becomes pure electric and longitudinal with
$\textbf{e}^{(2)}=\textbf{n}_\parallel$(and also non physical),
whereas $\textrm{a}^{(1)}_\mu$ is transverse
$\textbf{e}^{(1)}=\textbf{n}_\perp$, as it is $\textbf{e}^{(3)}$
\cite{shabad1,shabad2,shabad3}. In that case $\kappa_{(1)}=
\kappa_{(3)}$, and the circular polarization unit vectors
$(\textbf{e}^{{1}}\pm i \textbf{e}^{{3}})/\sqrt{2}$ are common
eigenvectors of $\Pi_{i j}$ and of the rotation generator matrix
$A^{3ij}$.

We want to remark three points: first, from the diagrammatic point
of view, the non-zero photon magnetic moment is due to the
non-vanishing three-leg diagram resulting from differentiating the
field -dependent photon self-energy with regard to $B$. Obviously,
such diagram vanishes in the limit $B =0$, as it is demanded by
Furry's theorem. As a second remark,  as $\mu_{\gamma}$  depends
through $\Pi_{\mu\nu}(k,k^{\prime\prime}\vert A^{ext})$ on the sums
over infinite pairs of Landau quantum numbers and spins of the
$e^{\pm}$ pairs, it cannot depend on any specific eigenvalue of
angular momentum, spin or orbit's center coordinates. Our third
remark is that due to the degeneracy of the orbit's centers of the
pair, a factor $eB$ is contained in
$\Pi_{\mu\nu}(k,k^{\prime\prime}\vert A^{ext})$ \cite{shabad2},
which is also even in $eB$ (as it is $\Omega_{EH}$), and
$\mu_{\gamma}$ is an odd function of $B$.

 The dispersion equations, obtained as the zeros of
the photon inverse Green function $D^{-1}_{\mu\nu}=0, $
 after diagonalizing the polarization operator
 $\Pi_{\mu\nu}(z_1,z_2,B)$, are
\begin{equation}
k^2=\kappa_{i}(z_2,z_1,B) \hspace{1cm} i=1,2,3.
\end{equation}
 After solving the dispersion equations for $z_1$ in terms of $z_2$ we
get
\begin{equation}
\omega^{(i)2}=\vert\textbf{k}\vert^2+\mathfrak{M}^{2(i)}\left(z_2,
B\right) \label{eg2}
\end{equation}

Let us remark at this point that the refraction index $n^{(i)}$ can
be defined as
\begin{equation}
n^{(i)}
=\frac{|\textbf{k}|}{\omega}=(1+\frac{\mathfrak{M}^{2(i)}}{\omega^2})^{1/2}
\label{refr}
\end{equation}
being different for each eigenmode, leading to the phenomenon of
birefringence. The propagation of light in magnetized vacuum is thus
similar to that in an anisotropic medium.  Gauge invariance
 implies that $\kappa_{(i)}(0,0, B)=0$ and $\mathfrak{M}^{2(i)}(0,B)=0$.
Thus, for parallel propagation $n^{(i)}=1$. By differentiating
(\ref{eg2}) with regard to $B$ we get the relation
$\mu_{\gamma}^{(i)}=-\frac{1}{2\omega}\frac{\partial
\mathfrak{M}^{2(i)}}{\partial B}$, and in consequence
$\mathfrak{M}^{2(i)}=-2\int_0^B \omega \mu_{\gamma}^{(i)}(B') dB' +
f(z_2)$. The function $f(z_2)$ is zero if the series expansion of
$\kappa_{(i)}$ in powers of $B$ is taken as linear in $z_1,z_2$ (see
below and Appendix). From (\ref{eg2}), we have the approximate
dispersion equation
\begin{equation}
\omega= |\textbf{k}| -\int_0^B \mu^{(i)}_{\gamma}
(z_2,B',|\textbf{k}|)dB'.
\end{equation}
For nonparallel propagation the fact that $n^{(i)}<1$ is a
consequence of photon paramagnetism.

Let us point out more explicitly the observable consequences of the
paramagnetic properties of the photon. As mentioned before, quantum
vacuum  has a magnetization ${\cal M}_V>0$ \cite{Elizabeth} at any
point of space $P$. If there is a nonzero photon density
$N_{\gamma}^i$ around $P$,  an additional magnetization ${\cal
M}_{\gamma}$, which in the monochromatic case it is simply ${\cal
M}_{\gamma}=\sum^i N_{\gamma}^i \mu_{\gamma}^i $, is produced on
$P$. This magnetization is essentially determined by the photon
momentum component $k_{\perp}$ (or angular momentum
$k_{\perp}/\sqrt{eB}$) transferred to the electron-positron virtual
quanta, and the photon gets its magnetic properties from its
interaction with the virtual pairs. The photon magnetization ${\cal
M}_{\gamma}$ contributes to an (usually small) increase of the
microscopic field $B$ to $B^{\prime}= B + 4\pi{\cal M}_{\gamma}$.

A possible way to measure the photon magnetic moment would be
provided by the measurement of the refraction indexes
$n^{(i)}$(\ref{refr}). Through them, we  obtain
$\mathfrak{M}^{2(i)}$, and by knowing $B$ and $\textbf{k}$, the
values of $\mu_{\gamma}^{(i)}$ would be obtained.  In the
astrophysical scenario, the dependence of $n^{(i)}$ with regard
$\mu_{\gamma}^{(i)}$ play an important role in the magnetic lensing
effect. See for instance \cite{Dupays}.

The renormalized eigenvalues of the polarization operator,
calculated in one-loop approximation \cite{shabad2}, are given by
the expressions
\begin{equation}\label{op-pol2}
\kappa_{i}=\frac{2 \alpha}{\pi}\int_{0}^{\infty}dt
\int_{-1}^{1}d\eta e^{-\frac{t}{b}
 }\left[\frac{\rho_{i}}{\sinh t}
e^{\zeta}+\frac{k^2 \bar{\eta}^2}{2t} \right],
\end{equation}
\begin{eqnarray*}
\zeta&=&-\frac{z_{2}}{eB}\frac{\sinh (\eta_+ t)\sinh (\eta_-
t)}{\sinh t}-\frac{z_1 }{eB}\bar{\eta}^2t\\
  \rho_1&=&-\frac{k^2}{2}\frac{\sinh (\eta_+ t)\cosh (\eta_+t)}{\sinh t}\eta_-
\\
  \rho_2&=&-\frac{z_1}{2}\bar{\eta}^2 \cosh t-\frac{z_2}{2}
  \frac{\sinh (\eta_+ t)\cosh (\eta_+t)}{\sinh t}\eta_ ,
\\
  \rho_3&=&-\frac{z_2}{2}\frac{\sinh (\eta_+ t)\sinh (\eta_- t)}
  {\sinh^{2} t} -\frac{z_1}{2}\frac{\sinh (\eta_+ t)\cosh (\eta_+
t)}{\sinh t} \eta_-,
\end{eqnarray*}
where we have used the notation $b=\frac{eB}{m^2}=\frac{B}{B_c}$,
$\eta_{\pm}=\frac{1 \pm \eta}{2}$, $\bar{\eta}=\sqrt{\eta_+\eta_-}$.

 \section{Photon anomalous magnetic moment}

We will find now the explicit expressions for the photon magnetic
moment. We can state that in the regions $-z_1,z_2 \leq 4m^2$ and
$0<B\leq B_c$, the photon is paramagnetic, since
$\mu_{\gamma}^{2,3}>0$. To that end we differentiate with regard to
$B$ the dispersion equation $z_{1}+z_{2}=\kappa_{i}$ and get
\begin{eqnarray} \label{FAMM1} \nonumber
   \frac{ \partial z_{1}}{\partial B} &=& \frac{\partial \kappa_{i}}{\partial
   B} \\
   &=& \frac{2 \alpha}{\pi}\int_{0}^{\infty}dt
\int_{-1}^{1}d\eta e^{-\frac{t}{b}}\left[ \frac{m^2}{B}\phi_{i}
+\frac{\partial z_{1}}{\partial B}\varphi_{i }\right],
\end{eqnarray}
\begin{eqnarray*}
\phi_{i}&=& \frac{1}{m^2}\left[\frac{\rho_{i}e^{\zeta}}{\sinh t}
\left(\frac{t}{b} -\zeta\right)+ \frac{k^2}{b}\frac{\bar{\eta}
^2}{2}\right] ,
\\  \varphi_{i }&=&
\frac{e^{\zeta}}{\sinh t} \left(\frac{\partial \rho_{i}}{\partial
z_{1}}-\frac{ \rho_{i}}{eB}\bar{\eta }^2t \right)+\frac{\bar{\eta
}^2}{2t},
\end{eqnarray*}
and taking in mind that $\frac{\partial z_{1}}{\partial B}=-2\omega
\frac{\partial \omega}{\partial B}$ in (\ref{FAMM1}), we obtain an
expression for the photon anomalous magnetic moment
\begin{eqnarray}\label{FAMM2}
    \mu_{\gamma}^i&=&-\frac{\partial \omega}{\partial B} \\
    \nonumber
    &=& \frac{m^2}{2 \omega B}\frac{\frac{2 \alpha}{\pi}\int_{0}^{\infty}dt
\int_{-1}^{1}d\eta e^{-\frac{t}{b} } \phi_{i}}{1-\frac{2
\alpha}{\pi}\int_{0}^{\infty}dt \int_{-1}^{1}d\eta e^{-\frac{t }{b}}
\varphi_{i }}.
\end{eqnarray}

By starting from the exact expressions (\ref{op-pol2}),
(\ref{FAMM2}) in Figs. 1,  2 it is depicted the result of a
numerical calculation for $\mu_{\gamma}$ for the second and third
modes, in the interval $0<B<B_c$, for frequencies such that $0<-z_1,
z_2 \simeq m^2$. It confirms that the paramagnetic behavior is
maintained throughout such interval for the modes $2,3$.
\begin{figure}[!htbp]
\includegraphics[width=4in]{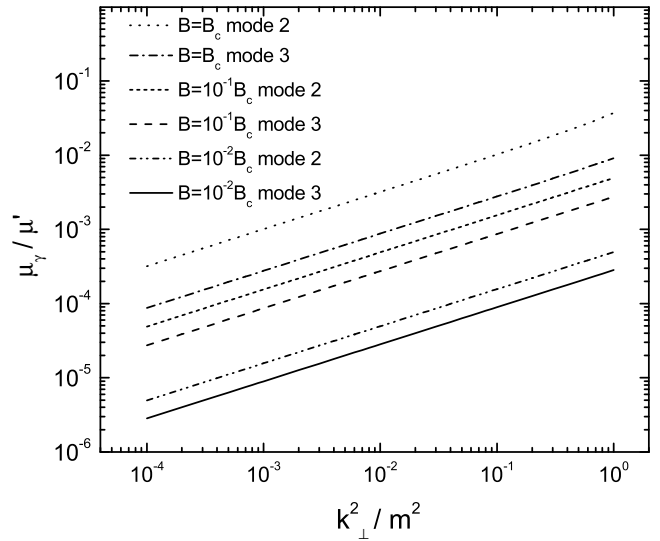}
\caption{\label{fig:mb2} Photon magnetic moment for modes 2 ,3 drawn
for $-z_1,z_2 \leq m^2$ in a logarithmic scale.}
\end{figure}
\begin{figure}[!htbp]
\includegraphics[width=3.7in]{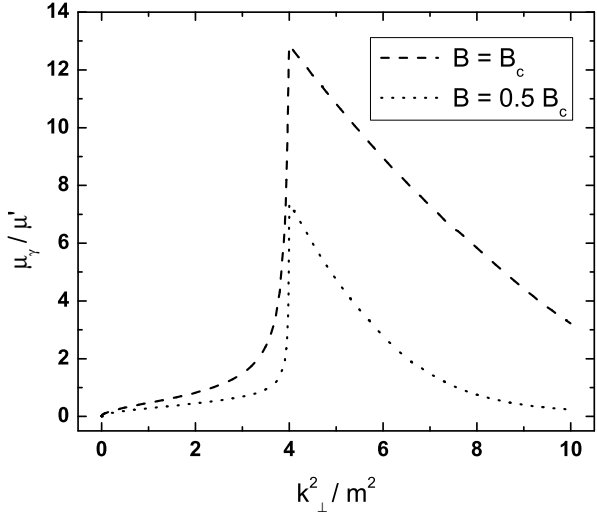}
\caption{\label{fig:Phvk} Photon magnetic moment curve for the
second mode drawn with regard to squared transverse momentum. It
shows a peak near the first threshold.}
\end{figure}

It is easy to see that for propagation along $B$ , the vacuum
behaves as in the limit $B=0$ for all eigenmodes. That is why we are
mainly interested in studying the perpendicular photon propagation
case $k_{\parallel}=0~$, for which the first mode is non physical,
as was pointed out before. As different from modes $2,3$ the
dispersion equation for mode $1$ (which is longitudinal and
unphysical for propagation orthogonal to $\textbf{B}$) is the light
cone in the low frequency limit $\omega^2_i-|\textbf{k}|^2=0$ (its
dependence on $B$ starts to appear in the term quadratic in $z_1,
z_2$; {see Appendix). We recall that this mode  exists as a physical
mode for propagation along $\textbf{B}$, which is on the light cone.

We consider the two extreme cases: 1) small departure from the light
cone dispersion law, and 2) large deviation from the light cone,
which occurs near the singularities of the photon self-energy due to
pair creation, which is especially interesting for fields $B\simeq
B_c$,  We discuss also the numerical calculation of
$\mu_{\gamma}^{2,3}$ in the region in between these two limits.

\subsection{Small frequencies and weak field limit}

 We will refer first to the  small frequencies and weak field case, which corresponds to
the subregions $-z_1,z_2 \ll m^2$ and  $0<B\ll B_c$. In that
subregion we can expand in series the expressions for $\kappa_{i}$
(\ref{op-pol2}), and retain only the linear approximation in $z_1,
z_2$ (the details of these calculations are given in the Appendix)
\begin{equation}\label{k0}
   \kappa_{i}^{(0)} = \frac{2 \alpha}{\pi} \sum_{j=1,2} G_{ij}z_{j},
    \end{equation}
where $G_{ij}(b)=\sum_{n=1}^{\infty} a_{ijn} b^{2n}$ are series in
even powers of $b$, and $a_{ijn}$ are numerical coefficients.

 In this section we are interested in
computing $\mu_{\gamma}^{2,3}$ (perpendicular propagation).
 The solutions of the dispersion equations in this limit
$z_1+z_2=\kappa_{2,3}^{(0)}$ for  the second and the third modes,
respectively, are
\begin{eqnarray} \label{dispeq2}
 z_1 &=& -z_2\frac{1- \frac{2 \alpha}{\pi}  G_{22}}{1-\frac{2 \alpha}{\pi} G_{21}} ,\\ \label{dispeq3}
 z_1 &=& -z_2 \frac{1-\frac{2 \alpha}{\pi}  G_{32}}{1-\frac{2 \alpha}{\pi} G_{31}}
\end{eqnarray}
 From (\ref{dispeq2}), (\ref{dispeq3}), after a straightforward calculation we get
$\mu_{\gamma}^{2,3}=-\frac{\partial \omega}{\partial B}$,

\begin{eqnarray}
 \nonumber
\mu_{\gamma}^{2} &=& \frac{\alpha z_2 }{\pi \omega B_c} \frac{\left(
1-\frac{2 \alpha}{\pi} G_{21}\right) \tilde{G}_{22}-\left(1- \frac{2
\alpha}{\pi} G_{22}\right)\tilde{G}_{21}}{\left( 1-\frac{2
\alpha}{\pi} G_{21} \right)^2},
\\
\nonumber
   \mu_{\gamma}^{3} &=& \frac{\alpha z_2 }{\pi \omega B_c} \frac{\left(
1-\frac{2 \alpha}{\pi} G_{31} \right)  \tilde{G}_{32}-\left(1-
\frac{2 \alpha}{\pi} G_{32}\right)\tilde{G}_{31}}{\left( 1-\frac{2
\alpha}{\pi} G_{31} \right)^2},
\\
 \tilde{G}_{ij}&=&\frac{d G_{ij}}{d b} = \frac{1}{b^{2}}
    \int_{0}^{\infty}dt
 e^{-\frac{t}{b} }
g_{ij}t.
\end{eqnarray}
 It is easy to see that $\mu_{\gamma}^{2,3}>0$ due to the smallness
of $\alpha <<1$ and also because $-\tilde{G}_{21}>\tilde{G}_{32}>
\tilde{G}_{22}>0$, $\tilde{G}_{22}=\tilde{G}_{31}$.

 As an example, we take the first two terms in the $\kappa_{i}^{(0)}$ series expansion. We can write $d^{(2)}=7\alpha/45 \pi$, $d^{(3)}=4\alpha/45 \pi$,
$c^{(2)}=26/49$, $c^{(3)}=-12/7$, (in the present approximation the
expressions for  $d^{(2),(3)}$ agree with those obtained earlier in
refs. \cite{Adler}, \cite{Dittrich} in calculating the indexes of
refraction of magnetized vacuum orthogonal and parallel to $B$, by
properly differentiating the Euler-Heisenberg Lagrangian). We have
then
\begin{equation}
\mathfrak{M}^{2(i)}\left(z_2, B\right)= - d^{(i)}z_2(b^2+c^{(i)}
b^4), \label{de0}
\end{equation}
and the magnetic moments
\begin{equation}
\mu_{\gamma}^{2}= \frac{14 \alpha z_2}{45 \pi B_c |\textbf{k}|}
\left(b -\frac{52 b^3}{49}\right)>0
\end{equation}
 and
\begin{equation}
 \mu_{\gamma}^{3}= \frac{8 \alpha z_2}{45 \pi B_c |\textbf{k}|}
\left(b -\frac{24 b^3}{7}\right)>0.
\end{equation}
In the present limit $ \mu_{\gamma}^{2,3}$ are small quantities
which grow with the frequency and magnetic field intensity. For
instance, for radiation of frequency $10^{15}$Hz and magnetic field
intensities of order $5\cdot 10^{4}$G (of order similar of those
used in some of the PVLAS experiments \cite{Zavattini}),
$\mu_{\gamma}^{2,3}\sim 10^{-16}\mu^{\prime}$. This quantity is very
small and a  photon density of order $N_{\gamma}\sim 10^{24}$
cm$^{-3}$, would be required to have the photon magnetization ${\cal
M}_{\gamma}^{2,3}=N_{\gamma}\mu_{\gamma}^{2,3}\sim 10^{-15}$G,
larger than the vacuum magnetization ${\cal M}_V$. But in magnetic
fields and densities of the same order but using X-ray lasers
\cite{laser} $\mu_{\gamma}^{2,3}\sim 10^{-13}\mu^{\prime}$, and the
magnetization might be of order $10^{-7}$G.

Larger values may occur for $\gamma $ rays in strongly magnetized
stars. For instance, for frequencies of order $10^{20}$Hz and
magnetic fields $\sim 10^{12}$G, we have $\mu_{\gamma}^{2,3}\sim
10^{-2}\mu^{\prime}$.

\subsection{High frequencies  and strong fields case  }

 We will be interested now in the case $m^2\lesssim -z_1\leq 4m^2 ,B\lesssim
B_c$. In such case, the energy gap between successive Landau energy
levels for electrons and positrons is of order close to the electron
rest energy. The photon self-energy diverges for values of  $-z_1=
k_{\perp}^{\prime 2}$, where
\begin{equation}
k_\perp^{\prime 2}=m^2[(1+2nb)^{1/2}+(1+2n^{\prime}b)^{1/2}]^2
\end{equation}
 are the thresholds for pair creation \cite{shabad1} at the  Landau quantum numbers $n,n'=0,1,2..$. Near
 and below the thresholds, the photon behaves like a massive vector particle. We are interested in the first threshold $n=n'=0$
  and photons of energy $\omega$ such that $-z_1 \lesssim 4m^2$.  The eigenvalues of the propagation modes can be written
 approximately near the thresholds $n,n^{\prime}$
\cite{Hugo2} as
\begin{equation}
\kappa_{(i)}\approx-2\pi\phi_{n,n^{\prime}}^{(i)}/\vert\Lambda\vert
\label{eg5}
\end{equation}
with
 \begin{equation}
\vert\Lambda\vert=((k_\perp^{\prime 2
}-k_\perp^{\prime \prime 2})(k_\perp^{\prime
2}-\omega^2+k_\parallel^2))^{1/2}
\end{equation}
 where
 \begin{equation}k_\perp^{\prime\prime
2}=m^2[(1+2nb)^{1/2}-(1+2n^{\prime}b)^{1/2}]^2,
\end{equation}
 is the
squared threshold energy for excitation between Landau levels
$n,n^{\prime}$ for an observable electron or positron. This term
does not lead to any singular behavior in the present quantum vacuum
case. The functions $\phi_{n,n^{\prime}}^{(i)}$ are expressed in
\cite{Hugo2} in terms of Laguerre functions of the variable $z_2/2e
B$. The expression (\ref{eg5}), being an approximation, is not even
in $B$ for a given pair $n, n^{\prime}$. An even expression would be
obtained after summing over all $n, n^{\prime}$ values.

In the vicinity of the first threshold $n=n^{\prime}=0$ and by
considering $k_\perp\neq0$ and $k_\parallel\neq0$, according to
\cite{shabad1,shabad2} the physical eigenwaves are described by the
second and third modes, but only the second mode has a singular
behavior near the threshold and the function $\phi^{(2)}_{0,0}$ has
the structure
\begin{equation}
\phi_{0,0}^{(2)}\simeq-\frac{2\alpha e B
m^2}{\pi}\textrm{exp}\left(-\frac{z_2}{2e B}\right),\label{00}
\end{equation}
In this case
 $k_{\perp}^{\prime\prime 2}=0$ and $k_{\perp}^{\prime2}=4m^2$ is the
 threshold energy. The solutions of the dispersion
equation $k^2=\kappa_{(2)}^{00}$ were obtained in \cite{shabad2}.
Below the thresholds it gives a real solution plus two complex
conjugate ones, located on the second sheet of the complex $\omega$
plane.

It is important to remark here the role of the function
$e^{-z_2/2eB}$ which is present in all $\phi_{n,n'}^{(i)}$. It makes
them significant for $z_2 \lesssim 2eB$ and makes them and
(\ref{eg5}) vanishing small for $z_2>>2eB$.
 By substituting (\ref{00}) in (\ref{eg5}), and by
differentiating the dispersion equation  with regard to $B$ one gets
for the second mode the photon  magnetic moment
\begin{equation}
\mu_\gamma^{(2)}=\frac{PQ}{\omega B_c
(P^{3/2}+bQ)}(1+\frac{z_2}{2eB})>0,\label{FRR1}
\end{equation}
\begin{equation}\nonumber
P=(4m^2+z_1), \hspace{.5cm} Q=\alpha m^3\exp(-\frac{z_2}{2eB}).
\end{equation}

 This solution is valid in the
vicinity of the first threshold and agrees with the previously
mentioned numerical result in its paramagnetic property, which is
valid throughout the whole region of transparency (below the first
pair creation threshold).

The expression (\ref{FRR1}) has a maximum near the threshold, $z_2
\simeq k_\perp^{\prime 2}$. If we consider $\omega$ close to $2m$,
the function $\mu_{\gamma}^{(2)}=f(X)$, where $X=\sqrt{4m^2 +z_1}$
has a maximum for $X= (2\pi\phi_{00}^{(2)}/m)^{1/3}$, which is very
close to the threshold. By calling
 \begin{equation}\label{mumax}
 m_{\gamma}=\omega (k_\perp^{\prime
2})=\sqrt{4m^2-m^2[2\alpha b\exp(-\frac{2}{b})]^{2/3}}
\end{equation}
 that maximum is $\mu_{\gamma}^{(2)}=\frac{e(1+2b)}{3m_\gamma b}\left[2\alpha
b\exp\left(-\frac{2}{b}\right)\right]^{2/3}$.
Numerically for $b
\sim 1$,  $ \mu_\gamma^{(2)}\approx 3\mu^\prime
\left(\frac{1}{2\alpha}\right)^{1/3}\approx 12.85\mu^\prime$.

 In Fig.1 it is
depicted the dependence of $\mu_\gamma^{(2)}$ with regard to
$k_\perp^{\prime 2}$. It shows that $\mu_\gamma^{(2)}>0$ and has an
absolute maximum  for $B\lesssim B_c$.

In (\ref{mumax}) we introduced $m_{\gamma}$ which has meaning near
the thresholds, and we name the \textit{photon dynamical mass} in
presence of a strong magnetic field. It is a consequence of the
enhanced coexistence of the massless photon with the massive virtual
pair (as a polariton). For energies near the thresholds, it has a
behavior similar to a neutral massive vector particle moving
parallel to $B$. However, it does not violate gauge invariance since
the condition $\Pi_{\mu \nu}(0,0,B) =0$ is preserved. The idea of a
photon mass has been introduced previously, for instance in
\cite{Osipov}, in a regime different from ours, in which
$k_{\parallel}\gg 4m^2$.

Beyond the first threshold $n=n'=0$, for frequencies such that $-z_1
>4m^2$ starts the so-called region of absorption, i.e., for
$4m^2+z_1<0$, $\kappa_{(2)}$ becomes complex, its imaginary part
leading to complex frequencies $\omega + i \Gamma$ after solving the
dispersion equations (the thresholds for absorption would be
slightly lower if the photon decays in a pair in a bound state,
forming positronium. See \cite{shabad3},\cite{Usov}). The quantity
$\Gamma$ is finite on the photon mass shell \cite{shabad2} and it
accounts for the probability of photon decay in electron-positron
pairs (the same occurs for higher thresholds). Thus, in this region
the photon magnetic moment cannot be considered independently of the
created electron-positron pairs.
\section{ Conclusions and Discussion}
We have seen that a photon moving in magnetized vacuum shows a
paramagnetic behavior, its magnetic moments, which are different for
each polarization mode, being increasing functions of $B$. This
behavior is shown in the low frequency, low magnetic field limit
$-z_1\ll m^2$, $B\ll B_c$ as well as in the large frequency large
magnetic field limit, $m^2 \leq -z_1\leq 4m^2$ $B\simeq B_c$. Both
limits correspond to the region of transparency. The measurement of
$\mu^{2,3}_{\gamma}$ in the low frequency limit could be accessible
to next future laboratory conditions. The large frequency limit is
interesting  in connection to astroparticle physics (strongly
magnetized stars).

The absorptive region is the continuation of the large frequency,
large magnetic field limit to the region  $-z_1\geq 4m^2$ and fields
$B \gtrsim B_c$. That region is also interesting in astrophysics,
and in cosmology (stars with fields $B \gtrsim B_c$  and early
universe). A study on the photon magnetic moment in this new
scenario is in progress by the present authors. It is interesting,
however, to remark some of the new features. For instance, although
larger values are expected for the photon magnetic moment than in
the region of transparency,  a negative peak is found for the first
threshold of the third mode. This has no absolute meaning since in
that region photons coexist with electron-positron pairs (the photon
has some nonzero probability of decaying  in pairs or either in
positronium) and the magnetic moment of the created
electron-positron pairs (or positronium) must be added to that of
photons. Also, the magnetized vacuum background is no longer
satisfactorily described by the Euler-Heisenberg expression
$\Omega_{EH}$, and radiative corrections containing the photon
self-energy must be taken into account. These corrections can be
written as $\Omega_{EH}^{1}= \sum_i \int_0^e (de'/e')\int d^{3}k
d\omega \kappa_{(i)}/(k^2-\kappa_{(i)})$ (or either, a similar
expression in terms of the electron Green function and self
energy)\cite{Fradkin} (appropriate counterterms must be subtracted
to make the integrals convergent). For some ranges of $k$ and
$\omega$, $\kappa_{(i)}$ become complex, and this suggests that
quantum vacuum modes at these  frequencies might become unstable and
decay for fields $B \gtrsim B_c$. At present QED in unstable vacuum
(see for instance \cite{Grib}, \cite{Fradkin2} and its references)
has growing interest, and the possibility of observing vacuum decay
in critical electric fields in terrestrial laboratories (see
\cite{Rafelski}) is becoming realistic thanks to the development of
high power pulse lasers technology.

\section{Acknowledgments}

The authors thank  A.E. Shabad, for several comments and important
remarks and to S. Villalba Chavez for several discussions. They
thank also OEA-ICTP for support under Net-35. H.P.R. thanks G.
Altarelli and J. Ellis for comments, and to CERN for hospitality at
early stages of this research.

\section{Appendix}

We start from the renormalized eigenvalues of the polarization
operator in presence of a constant homogeneous magnetic field in the
one-loop approximation, given by Shabad (\ref{op-pol2}). We are
interested in a wide range of frequencies characterized by the
condition $z_1 ,z_2<<m^2$. We can express $\kappa_{i}$ as

\begin{equation}\label{series-op-pol}
 \kappa_{i}=\sum_{l=0}^{\infty}
\kappa_{i}^{(l)},
 \end{equation}
\begin{eqnarray*}
   \kappa_{i}^{(0)} &=&
\frac{2 \alpha}{\pi } \int_{0}^{\infty}dt \int_{-1}^{1}d\eta
e^{-\frac{t}{b} } \left[\frac{\rho_{i}}{\sinh  t} +(z_1
+z_2)\frac{1-\eta ^2}{8t} \right], \\
\kappa_{i}^{(l)} &=& \frac{2 \alpha}{\pi } \frac{1}{l!}
\int_{0}^{\infty}dt \int_{-1}^{1}d\eta e^{-\frac{t}{b}}
\frac{\rho_{i}}{\sinh t} \zeta^{l},
\end{eqnarray*}
and consider  the first few terms of $\kappa_{i}^{(l)}$ in the
series expansion (\ref{series-op-pol}). Here we explicitly compute
$\kappa_{i}^{(0)}$ and $\kappa_{i}^{(1)}$.
\subsection{The linear in $z_1, z_2$ term $\kappa_{i}^{(0)}$}

After integrating on $\eta$ we get for the term $\kappa_{i}^{(0)}$ (
$i=1,2,3$), which is linear in $z_1, z_2$
\begin{equation}\label{k0}
   \kappa_{i}^{(0)} =  \frac{ 2\alpha}{\pi}
    \int_{0}^{\infty}dt
 e^{-\frac{t}{b} } \sum_{j=1,2}
g_{ij}z_{j},
\end{equation}
\begin{eqnarray}
     g_{11} &=& -\frac{1}{12\,t } - \frac{\coth t }{4\,{t}^2} +
\frac{\coth^2 t }{4\,t }>0, \\ \nonumber
   g_{21} &=&\frac{1}{6t} - \frac{\cosh t}{6\sinh t}<0, \\ \nonumber
   g_{32}  &=& \frac{1}{6 t }+\frac{1}{2 t
\sinh^2 t} -\frac{\cosh t}{2 \sinh^3 t} >0,
\end{eqnarray}
and \begin{equation}
    g_{12} = g_{22}  = g_{31}  =g_{11}.\label{equg}
\end{equation}

For fields $eB<<m^2$ (actually it is enough that $\frac{eB}{m^2}
\lesssim 10^{-1}$) the functions inside the integrals in
$\kappa_{i}^{(0)}$ are significantly different from zero only for
$t<<1$ and we can expand the following expressions  around $t=0$ and
retain the first four terms:

\begin{eqnarray} \label{g} \nonumber
g_{11}&\approx& \frac{t }{45} - \frac{{t }^3}{315} +
\frac{2\,{t}^5}{4725} - \frac{{t }^7}{18711}
,\\
  g_{21} &\approx& \frac{1}{6 } \left(-\frac{t}{3}+\frac{t^3}{45}
  -\frac{2 t^5}{945}+\frac{t^7}{4725}\right),\\ \nonumber
    g_{32}&\approx&
\frac{t}{15}-\frac{t^3}{63}
  +\frac{2 t^5}{675}-\frac{t^7}{2079},
\end{eqnarray}

Finally, by using (\ref{g}) we easily get approximate expressions
for the first term $\kappa_{i}^{(0)}$ of the series expansion of the
eigenvalues (\ref{series-op-pol}). We have

\begin{equation}\label{k0}
   \kappa_{i}^{(0)} = \frac{2 \alpha}{\pi} \sum_{j=1,2} G_{ij}z_{j},
    \end{equation}
where
    \begin{equation}\label{G}
   G_{ij} =
    \int_{0}^{\infty}dt
 e^{-\frac{t}{b} }
g_{ij},
    \end{equation}
are even functions of $b$ (from (\ref{equg})
$G_{11}=G_{12}=G_{22}=G_{31})$
\begin{eqnarray*}
G_{11}&=& \frac{1}{45}b^2 - \frac{2}{105}b^4
    + \frac{16}{315}b^6
   - \frac{80}{297}b^8,
\\
    G_{21}&=& \frac{1}{6}
\left[-\frac{1}{3}b^2 +\frac{2}{15} b^4-\frac{16}{63}b^6
+\frac{16}{15}b^8\right] ,\\
    G_{32}&=&
 \frac{1}{15}b^2
-\frac{2}{21}b^4+\frac{16}{45}b^6 -\frac{80}{33}b^8,
\end{eqnarray*}

\subsection{The term $\kappa_{i}^{(1)}$, quadratic  in $z_1,z_2$}
The  quadratic terms in $z_1, z_2$ appear in the term
$\kappa_{i}^{(1)}$, which can be calculated by following the same
 procedure we used previously  to get $\kappa_{i}^{(0)}$. We obtain, after
 integrating in $\eta$,
\begin{equation}
    \kappa_{i}^{(1)} =  \frac{\alpha}{\pi  eB}
\int_{0}^{\infty}dt
  e^{-\frac{m^2}{eB}t } \sum_{j,n=1,2} f_{ijn}z_{j}z_{n},
  \end{equation}
   \begin{eqnarray*}
    \nonumber f_{111}&=& -\frac{1}{4\,t ^2} - \frac{3\,\coth t }{4\,t^3}
   + \frac{3\,{\coth^2 t}}{4\,t^2}, \\
   \nonumber
   f_{112} &=& \frac{1}{2\sinh^2 t }\left[-\frac{1}{8 }+\frac{19}{32 t^2}
   +\frac{11 \cosh 2 t}{32 t^2 } - \right.  \\ \nonumber
   & & \hspace{3.5cm}\left.\frac{3 \coth  t}{16 t }-\frac{3 \sinh 2 t}{8
   t^3}\right],
 \\
\nonumber
 f_{122} &=& \frac{3 \coth^2 t}{16 t^2} -\frac{1}{8 \sinh^2 t}
  -\frac{3 \coth t}{16 t \sinh^2 t},\\
\nonumber f_{211} &=& \frac{t}{15}\coth t ,\\
 \nonumber
 f_{212} &=&\frac{1}{24}\left[2-\frac{3\coth t}{ t
  ^{3}}+\left(2+\frac{3}{t^2}\right)\frac{1}{\sinh^2 t}\right],
 \\
\nonumber
 f_{222}  &=& \frac{1}{16 t^2} \left[3\coth^2 t-
  t\left(2 t +3\coth t \right)\frac{1}{\sinh^2 t}\right],\\
 \nonumber
 f_{311} &=& -\frac{1}{4 t^3} \left(t+3\coth t- 3t\coth^2 t\right),\\
\nonumber
 f_{312} &=&\frac{1}{96t^2}\left[9+
  \frac{33-6t^2+t(-33+8t^2)\coth
t }{\sinh^2 t}\right],
 \\
\nonumber
 f_{322} &=& \frac{1}{8\sinh^4t} \left[4+2\cosh(2 t)-
  \frac{3}{t}\sinh (2t) \right].
  \end{eqnarray*}

   After expanding the
  functions $f_{ijn}$ around $t=0$, by keeping  the first five terms in
  the expansion, we have
\begin{eqnarray} \nonumber
  f_{111}
   &\approx&\frac{1}{15} - \frac{t ^2}{105} + \frac{2\,t^4}{1575} - \frac{t^6}{6237} +
  \frac{1382\,t^8}{70945875} ,\\ \nonumber
   f_{112} &\approx& \frac{1}{2}\left[\frac{2}{15} - \frac{t^2}{42} + \frac{t^4}{270} - \frac{82\,t^6}{155925} +
  \frac{907\,t ^8}{12899250}\right] ,
  \\ \nonumber
  f_{122}
  &\approx& \frac{1}{15} - \frac{t ^2}{70} + \frac{23\,t ^4}{9450} - \frac{19\,t ^6}{51975} +
  \frac{7213\,t^8}{141891750} ,
\\
  f_{211} &\approx&  \frac{1}{15}\left(1+\frac{t^2}{3}
  -\frac{t^4}{45}+\frac{2t^6}{945}-\frac{t^8}{4725}\right),\\ \nonumber
  f_{212} &\approx& \frac{1}{2}\left[\frac{2}{15}+\frac{t^2}{126}-\frac{19t^4}{14175}
  +\frac{181 t^6}{935550}-\frac{5443t^8}{212837625}\right],
  \end{eqnarray}
\begin{eqnarray}\nonumber
   f_{222}
  &\approx& \frac{1}{15}-\frac{t^2}{70}+\frac{23t^4}{9450}
  -\frac{19 t^6}{51975}+\frac{7213t^8}{141891750},
\\ \nonumber
  f_{311}
   &\approx&\frac{1}{15}-\frac{t^2}{105}
  +\frac{2t^4}{1575}-\frac{t^6}{6237}+\frac{1382t^8}{70945875},\\ \nonumber
  f_{312} &\approx& \frac{1}{2}\left[\frac{2}{15}-\frac{13t^2}{315}+\frac{253t^4}{28350}
  -\frac{ t^6}{630}-\frac{106643t^8}{425675250}\right],
  \\ \label{k1} \nonumber
  f_{322}
  &\approx& \frac{1}{15}-\frac{2t^2}{63}+\frac{2t^4}{225}
  -\frac{4 t^6}{2079}+\frac{1382t^8}{3869775}.
\end{eqnarray}

We finally integrate once more in $t$ and express $\kappa_{i}^{(1)}$
as
\begin{equation}\label{k1}
   \kappa_{i}^{(1)} = \frac{\alpha}{\pi  m^2} \sum_{j,n=1,2}
   F_{ijn}z_{j}z_{n},
    \end{equation}
where
  \begin{equation}\label{F}
  F_{ijl} =\frac{1}{b} \int_{0}^{\infty}dt
 e^{-\frac{t}{b}}f_{ijn},
\end{equation}
\begin{eqnarray*}
    F_{111}&=& \frac{1}{15}- \frac{2}{105}b^2 +
    \frac{16}{525}b^4- \frac{80}{693}b^6
      +\frac{176896}{225225}b^8,
\\
F_{112}&=& \frac{1}{2}\left[\frac{2}{15}-
  \frac{1}{21}b^2+ \frac{4}{45}b^4
   - \frac{1312}{3465}b^6+ \frac{58048}{20475}b^8\right],
   \\
 F_{122}&=&  \frac{1}{15}-
  \frac{1}{35}b^2 + \frac{92}{1575}b^4
  - \frac{304}{1155}b^6+ \frac{461632}{225225}b^8,
  \\
    F_{211}&=&\frac{1}{15}\left[1+\frac{2}{3}
b^2 -\frac{8}{15}b^4+\frac{32}{21}b^6 -\frac{128}{15}b^8\right],
\\
 F_{212}&=&\frac{1}{2}\left[\frac{2}{15}+\frac{1}{63}b^2
-\frac{152}{4725}b^4+\frac{1448}{10395}b^6 -\frac{696704}{675675}b^8
\right],
\\
 F_{222}&=&\frac{1}{15}-\frac{1}{35}b^2
+\frac{92}{1575}b^4-\frac{304}{1155}b^6 +\frac{461632}{225225}b^8,
\\
    F_{311}&=&\frac{1}{15}-\frac{2}{105}
b^2 +\frac{16}{525}b^4-\frac{80}{693}b^6 +\frac{176896}{225225}b^8,
\\  F_{312}&=&
 \frac{1}{2}\left[\frac{2}{15}-\frac{26}{315}b^2
+\frac{1012}{4725}b^4-\frac{8}{7}b^6
+\frac{6825152}{675675}b^8 \right], \\
 F_{322} &=& \frac{1}{15}-\frac{4}{63}b^2
+\frac{16}{75}b^4-\frac{320}{231}b^6 +\frac{176896}{12285}b^8.
\end{eqnarray*}
Note that in the zero field limit $B=0$, the term
$\kappa_{i}^{(1)}$, as different from $\kappa_{i}^{(0 }$, does not
vanish, and contains the factor $(z_1 + z_2)^2$. This guarantees
that the photon obeys the light cone dispersion law $z_1 +
z_2=k^2=0$ in absence of the external field $B$ \cite{Fradkin}.

\bibliographystyle{apsrev}

\end{document}